\pdfoutput=1

\documentclass[12pt,a4paper]{article}

\usepackage{ifthen} 
\newboolean{usetwocolumn}
\setboolean{usetwocolumn}{false}

\newboolean{pdflatex}
\setboolean{pdflatex}{true} 

\newboolean{articletitles}
\setboolean{articletitles}{true} 

\newboolean{uprightparticles}
\setboolean{uprightparticles}{false} 

\newboolean{inbibliography}
\setboolean{inbibliography}{false} 
\def\paperauthors{LHCb collaboration} 

\def\papercopyright{CERN on behalf of the LHCb collaboration}
\def\paperasciititle{Measurement of the ratio of branching fractions
  B(B_c^+ -> J/psi tau^+ nu_tau)/B(B_c^+ -> J/psi mu^+ nu_mu)}
\def\papertitle{Measurement of the ratio of branching fractions $\mathcal{B}(\Bc\,\to\,\jpsi\taup\neut)$/$\mathcal{B}(\Bc\,\to\,\jpsi\mup\neum)$}
\def\paperkeywords{{High Energy Physics}, {LHCb}} 
\def\papercopyright{CERN on behalf of the LHCb collaboration}
\def\paperlicence{CC-BY-4.0}
\def\paperlicenceurl{https://creativecommons.org/licenses/by/4.0/}

\usepackage[top=1in, bottom=1.25in, left=1in, right=1in]{geometry}

\usepackage{microtype}
\usepackage{lineno}  
\usepackage{xspace} 
\usepackage{caption} 

\usepackage{graphicx}  
\usepackage{color}
\usepackage{colortbl}
\graphicspath{{./figs/}} 

\usepackage{amsmath} 
\usepackage{amssymb}
\usepackage{amsfonts}
\usepackage{upgreek} 

\newcommand*\patchAmsMathEnvironmentForLineno[1]{%
\expandafter\let\csname old#1\expandafter\endcsname\csname #1\endcsname
\expandafter\let\csname oldend#1\expandafter\endcsname\csname
end#1\endcsname
 \renewenvironment{#1}%
   {\linenomath\csname old#1\endcsname}%
   {\csname oldend#1\endcsname\endlinenomath}%
}
\newcommand*\patchBothAmsMathEnvironmentsForLineno[1]{%
  \patchAmsMathEnvironmentForLineno{#1}%
  \patchAmsMathEnvironmentForLineno{#1*}%
}
\AtBeginDocument{%
\patchBothAmsMathEnvironmentsForLineno{equation}%
\patchBothAmsMathEnvironmentsForLineno{align}%
\patchBothAmsMathEnvironmentsForLineno{flalign}%
\patchBothAmsMathEnvironmentsForLineno{alignat}%
\patchBothAmsMathEnvironmentsForLineno{gather}%
\patchBothAmsMathEnvironmentsForLineno{multline}%
\patchBothAmsMathEnvironmentsForLineno{eqnarray}%
}

\usepackage{hyperxmp}

\usepackage[pdftex,
            pdfauthor={\paperauthors},
            pdftitle={\paperasciititle},
            pdfkeywords={\paperkeywords},
            pdfcopyright={Copyright (C) \papercopyright},
            pdflicenseurl={\paperlicenceurl}]{hyperref}

\usepackage[all]{hypcap} 

\usepackage{xspace} 
\usepackage{upgreek}

\def\lhcb {\mbox{LHCb}\xspace}

\def\MagUp {\mbox{\em Mag\kern -0.05em Up}\xspace}

\ifthenelse{\boolean{uprightparticles}}%
{

 \def\Pmu         {\ensuremath{\upmu}\xspace}                 
 \def\Pnu         {\ensuremath{\upnu}\xspace}                 
                  
 \def\Ppi         {\ensuremath{\uppi}\xspace}

 \def\Ptau        {\ensuremath{\uptau}\xspace}

 \def\Pchi        {\ensuremath{\upchi}\xspace}                 
 \def\Ppsi        {\ensuremath{\uppsi}\xspace}

 \def\PDelta      {\ensuremath{\Delta}\xspace}                 
 \def\PXi      {\ensuremath{\Xi}\xspace}                 
 \def\PLambda      {\ensuremath{\Lambda}\xspace}                 
 \def\PSigma      {\ensuremath{\Sigma}\xspace}                 
 \def\POmega      {\ensuremath{\Omega}\xspace}                 
 \def\PUpsilon      {\ensuremath{\Upsilon}\xspace}                 
 
 \def\PB      {\ensuremath{\mathrm{B}}\xspace}                 
                  
 \def\PD      {\ensuremath{\mathrm{D}}\xspace}

 \def\PJ      {\ensuremath{\mathrm{J}}\xspace}                 
 \def\PK      {\ensuremath{\mathrm{K}}\xspace}

 \def\Pb      {\ensuremath{\mathrm{b}}\xspace}                 
 \def\Pc      {\ensuremath{\mathrm{c}}\xspace}

 \def\Pi      {\ensuremath{\mathrm{i}}\xspace}

 \def\Ps      {\ensuremath{\mathrm{s}}\xspace}

}
{

 \def\Pmu         {\ensuremath{\mu}\xspace}                 
 \def\Pnu         {\ensuremath{\nu}\xspace}                 
                  
 \def\Ppi         {\ensuremath{\pi}\xspace}

 \def\Ptau        {\ensuremath{\tau}\xspace}

 \def\Pchi        {\ensuremath{\chi}\xspace}                 
 \def\Ppsi        {\ensuremath{\psi}\xspace}                 
                  
 \mathchardef\PDelta="7101
 \mathchardef\PXi="7104
 \mathchardef\PLambda="7103
 \mathchardef\PSigma="7106
 \mathchardef\POmega="710A
 \mathchardef\PUpsilon="7107
                  
 \def\PB      {\ensuremath{B}\xspace}                 
                  
 \def\PD      {\ensuremath{D}\xspace}

 \def\PJ      {\ensuremath{J}\xspace}                 
 \def\PK      {\ensuremath{K}\xspace}

 \def\Pb      {\ensuremath{b}\xspace}                 
 \def\Pc      {\ensuremath{c}\xspace}

 \def\Pi      {\ensuremath{i}\xspace}

 \def\Ps      {\ensuremath{s}\xspace}

}

\makeatletter
\ifcase \@ptsize \relax
  \newcommand{\miniscule}{\@setfontsize\miniscule{4}{5}}
\or
  \newcommand{\miniscule}{\@setfontsize\miniscule{5}{6}}
\or
  \newcommand{\miniscule}{\@setfontsize\miniscule{5}{6}}
\fi
\makeatother

\DeclareRobustCommand{\optbar}[1]{\shortstack{{\miniscule (\rule[.5ex]{1.25em}{.18mm})}
  \\ [-.7ex] $#1$}}

\def\mup        {{\ensuremath{\Pmu^+}}\xspace}
\def\mun        {{\ensuremath{\Pmu^-}}\xspace} 

\def\taup       {{\ensuremath{\Ptau^+}}\xspace}
\def\taum       {{\ensuremath{\Ptau^-}}\xspace}

\def\neu        {{\ensuremath{\Pnu}}\xspace}
\def\neub       {{\ensuremath{\overline{\Pnu}}}\xspace}
\def\neum       {{\ensuremath{\neu_\mu}}\xspace}
\def\neumb      {{\ensuremath{\neub_\mu}}\xspace}
\def\neut       {{\ensuremath{\neu_\tau}}\xspace}
\def\neutb      {{\ensuremath{\neub_\tau}}\xspace}

\def\squark    {{\ensuremath{\Ps}}\xspace}

\def\cquark    {{\ensuremath{\Pc}}\xspace}

\def\bquark    {{\ensuremath{\Pb}}\xspace}

\def\pion   {{\ensuremath{\Ppi}}\xspace}

\def\pip    {{\ensuremath{\pion^+}}\xspace}
\def\pim    {{\ensuremath{\pion^-}}\xspace}

\def\kaon    {{\ensuremath{\PK}}\xspace}
  \def\Kbar    {{\kern 0.2em\overline{\kern -0.2em \PK}{}}\xspace}

\def\KorKbar    {\kern 0.18em\optbar{\kern -0.18em K}{}\xspace}

\def\Kp      {{\ensuremath{\kaon^+}}\xspace}

  \def\Dbar    {{\kern 0.2em\overline{\kern -0.2em \PD}{}}\xspace}
\def\D       {{\ensuremath{\PD}}\xspace}

\def\DorDbar    {\kern 0.18em\optbar{\kern -0.18em D}{}\xspace}

\def\Dstar   {{\ensuremath{\D^*}}\xspace}

\def\Ds      {{\ensuremath{\D^+_\squark}}\xspace}

\def\B       {{\ensuremath{\PB}}\xspace}
\def\Bbar    {{\ensuremath{\kern 0.18em\overline{\kern -0.18em \PB}{}}}\xspace}

\def\BorBbar    {\kern 0.18em\optbar{\kern -0.18em B}{}\xspace}

\def\Bc      {{\ensuremath{\B_\cquark^+}}\xspace}
\def\Bcp     {{\ensuremath{\B_\cquark^+}}\xspace}

\def\jpsi     {{\ensuremath{{\PJ\mskip -3mu/\mskip -2mu\Ppsi\mskip 2mu}}}\xspace}
\def\psitwos  {{\ensuremath{\Ppsi{(2S)}}}\xspace}

\def\chiczero {{\ensuremath{\Pchi_{\cquark 0}}}\xspace}
\def\chicone  {{\ensuremath{\Pchi_{\cquark 1}}}\xspace}
\def\chictwo  {{\ensuremath{\Pchi_{\cquark 2}}}\xspace}
  \def\Y#1S{\ensuremath{\PUpsilon{(#1S)}}\xspace}

\def\chic  {{\ensuremath{\Pchi_{c}}}\xspace}

\def\Lbar        {{\ensuremath{\kern 0.1em\overline{\kern -0.1em\PLambda}}}\xspace}
\def\LorLbar    {\kern 0.18em\optbar{\kern -0.18em \PLambda}{}\xspace}

\newcommand{\decay}[2]{\ensuremath{#1\!\to #2}\xspace}         

\def\to                 {\ensuremath{\rightarrow}\xspace}

\def\AT#1     {\ensuremath{A_{\mathrm{T}}^{#1}}\xspace}           

\def\C#1      {\ensuremath{\mathcal{C}_{#1}}\xspace}                       
\def\Cp#1     {\ensuremath{\mathcal{C}_{#1}^{'}}\xspace}                    
\def\Ceff#1   {\ensuremath{\mathcal{C}_{#1}^{\mathrm{(eff)}}}\xspace}        
\def\Cpeff#1  {\ensuremath{\mathcal{C}_{#1}^{'\mathrm{(eff)}}}\xspace}       
\def\Ope#1    {\ensuremath{\mathcal{O}_{#1}}\xspace}                       
\def\Opep#1   {\ensuremath{\mathcal{O}_{#1}^{'}}\xspace}                    

\newcommand{\tev}{\ifthenelse{\boolean{inbibliography}}{\ensuremath{~T\kern -0.05em eV}}{\ensuremath{\mathrm{\,Te\kern -0.1em V}}}\xspace}
\newcommand{\gev}{\ensuremath{\mathrm{\,Ge\kern -0.1em V}}\xspace}
\newcommand{\mev}{\ensuremath{\mathrm{\,Me\kern -0.1em V}}\xspace}
\newcommand{\kev}{\ensuremath{\mathrm{\,ke\kern -0.1em V}}\xspace}
\newcommand{\ev}{\ensuremath{\mathrm{\,e\kern -0.1em V}}\xspace}
\newcommand{\gevc}{\ensuremath{{\mathrm{\,Ge\kern -0.1em V\!/}c}}\xspace}
\newcommand{\mevc}{\ensuremath{{\mathrm{\,Me\kern -0.1em V\!/}c}}\xspace}
\newcommand{\gevcc}{\ensuremath{{\mathrm{\,Ge\kern -0.1em V\!/}c^2}}\xspace}
\newcommand{\gevgevcccc}{\ensuremath{{\mathrm{\,Ge\kern -0.1em V^2\!/}c^4}}\xspace}
\newcommand{\mevcc}{\ensuremath{{\mathrm{\,Me\kern -0.1em V\!/}c^2}}\xspace}

\def\invfb   {\ensuremath{\mbox{\,fb}^{-1}}\xspace}

\newcommand{\stat}{\ensuremath{\mathrm{\,(stat)}}\xspace}
\newcommand{\syst}{\ensuremath{\mathrm{\,(syst)}}\xspace}

\def\gsim{{~\raise.15em\hbox{$>$}\kern-.85em
          \lower.35em\hbox{$\sim$}~}\xspace}
\def\lsim{{~\raise.15em\hbox{$<$}\kern-.85em
          \lower.35em\hbox{$\sim$}~}\xspace}

\def\pt         {\mbox{$p_{\mathrm{ T}}$}\xspace}

\def\tell1  {TELL1\xspace}
\def\ukl1   {UKL1\xspace}

\newcommand{\eg}{\mbox{\itshape e.g.}\xspace}
\newcommand{\ie}{\mbox{\itshape i.e.}\xspace}

\def\RDst {\ensuremath{\mathcal{R}(\Dstar)}\xspace}

\def\RD {\mathcal{R}(D)}
\def\mmsq {\ensuremath{m^2_{\rm miss}}\xspace}

\def\bctaunu {\decay{\Bc}{\jpsi \taup \neut}}
\def\bcmunu {\decay{\Bc}{\jpsi \mup \neum}}

\usepackage[separate-uncertainty=true]{siunitx}
\usepackage{booktabs}

\def\mmsq{\ensuremath{m_\mathrm{miss}^2}}
\def\emu{\ensuremath{E^*_\mu}}
\def\rjpsi{\ensuremath{\mathcal{R}(\jpsi)}}
\def\jpsimu{\ensuremath{\jpsi\mup}}

\usepackage[percent]{overpic}
\usepackage{subcaption}
\usepackage{multicol}

\usepackage{cite} 
\usepackage{mciteplus}
\begin{document}


\renewcommand{\thefootnote}{\fnsymbol{footnote}}
\setcounter{footnote}{1}

\ifthenelse{\boolean{usetwocolumn}}{\onecolumn}{}

\begin{titlepage}
\pagenumbering{roman}

\vspace*{-1.5cm}
\centerline{\large EUROPEAN ORGANIZATION FOR NUCLEAR RESEARCH (CERN)}
\vspace*{1.5cm}
\noindent
\begin{tabular*}{\linewidth}{lc@{\extracolsep{\fill}}r@{\extracolsep{0pt}}}
\ifthenelse{\boolean{pdflatex}}
{\vspace*{-2.7cm}\mbox{\!\!\!\includegraphics[width=.14\textwidth]{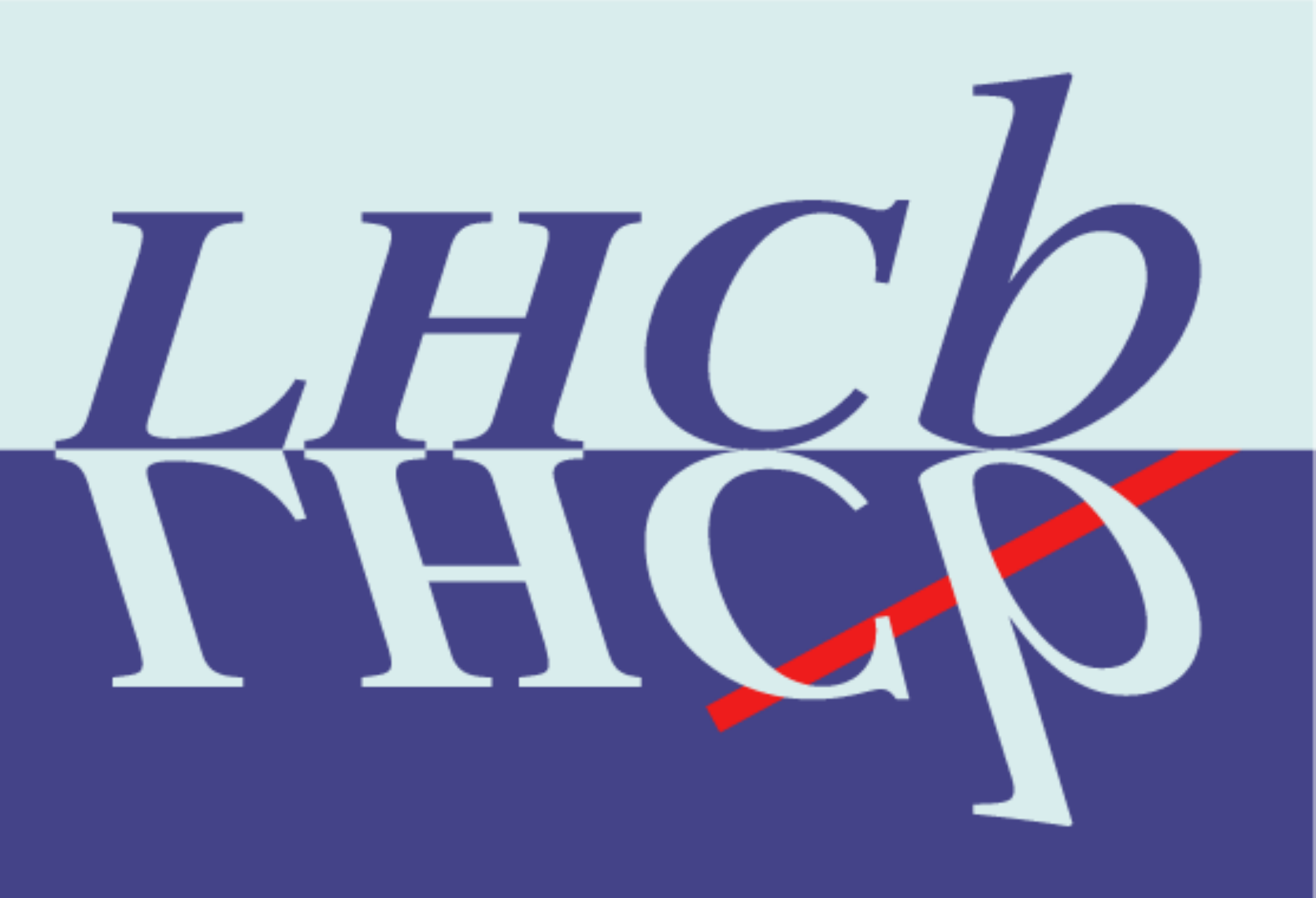}} & &}%
{\vspace*{-1.2cm}\mbox{\!\!\!\includegraphics[width=.12\textwidth]{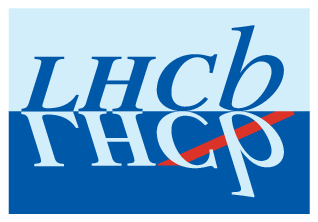}} & &}%
\\
 & & CERN-EP-2017-275\\  
 & & LHCb-PAPER-2017-035 \\  
 & & March 23, 2018 \\ 
 & & \\
\end{tabular*}

\vspace*{4.0cm}

{\normalfont\bfseries\boldmath\LARGE
\begin{center}
  \papertitle 
\end{center}
}

\vspace*{2.0cm}

\begin{center}
\paperauthors\footnote{Authors are listed at the end of this letter.}
\end{center}

\vspace{\fill}

\begin{abstract}
  A measurement is reported of the ratio of branching fractions $\rjpsi=\mathcal{B}(\Bc\,\to\,\jpsi\taup\neut)/\mathcal{B}(\Bc\,\to\,\jpsi \mup\neum)$, where the $\taup$ lepton is identified in the decay mode $\taup\,\to\,\mup\neum\neutb$.
 This analysis uses a sample of proton-proton collision data corresponding to 3.0\invfb of integrated luminosity recorded with the LHCb experiment at center-of-mass energies $7\tev$ and $8\tev$.
 A signal is found for the decay $\Bc\,\to\,\jpsi \taup\neut$ at a significance of 3 standard deviations, corrected for systematic uncertainty, and the ratio of the branching fractions is measured to be  $\rjpsi = 0.71 \pm 0.17 \stat\,  \pm 0.18\syst$.
 This result lies within 2 standard deviations above the range of central values currently predicted from the Standard Model. 
\end{abstract}

\vspace*{2.0cm}

\begin{center}
Published in Phys.~Rev.~Lett. 120, 121801 (2018) \end{center}

\vspace{\fill}

{\footnotesize 
\centerline{\copyright~\papercopyright, licence \href{\paperlicenceurl}{\paperlicence}.}}
\vspace*{2mm}

\end{titlepage}


\newpage
\setcounter{page}{2}
\mbox{~}
%
%
%
%

\cleardoublepage

\ifthenelse{\boolean{usetwocolumn}}{\twocolumn}{}

\renewcommand{\thefootnote}{\arabic{footnote}}
\setcounter{footnote}{0}



\pagestyle{plain} 
\setcounter{page}{1}
\pagenumbering{arabic}

Semileptonic \bquark-hadron decays provide powerful probes for testing the Standard Model (SM) and for searching for the effects of physics beyond the SM. 
Due to their relatively simple theoretical description via tree-level processes in the SM, these decay modes serve as an 
ideal setting for examining the universality of the couplings of the three charged leptons in electroweak interactions.  Recent measurements of the parameters $\RD$ and $\RDst$, corresponding to the ratios of branching fractions
$\mathcal{B}(\B\,\to\,D^{(*)}\taum\neutb)/\mathcal{B}(\B\to D^{(*)}\mun\neumb)$,
by the BaBar\cite{Lees:2012xj,Lees:2013uzd}, Belle\cite{Huschle:2015rga, Sato:2016svk,Hirose:2016wfn, Hirose:2017dxl} and LHCb\cite{LHCb-PAPER-2015-025, LHCb-PAPER-2017-017, LHCb-PAPER-2017-027} collaborations indicate larger values than the SM predictions~\cite{HFLAV}.
Proposed explanations for these discrepancies include extensions of the SM that involve enhanced weak couplings to third-generation leptons and quarks, such as interactions involving a
charged Higgs boson~\cite{Tanaka:1994ay,Crivellin:2012ye}, leptoquarks~\cite{freytsis:2015qca}, or new vector bosons~\cite{crivellin:2015lwa}.
Furthermore, other hints of the failure of lepton flavor universality have been seen in electroweak loop-induced \B-meson decays \cite{LHCb-PAPER-2014-024,LHCb-PAPER-2017-013}.

Measurements of semitauonic decays of other species of \bquark\ hadrons can provide additional handles for investigating the sources of theoretical and experimental uncertainties, and potentially the origin of lepton nonuniversal couplings.  
This Letter presents the first study of the semitauonic decay $\bctaunu$  and a measurement of the ratio of branching fractions
\begin{equation}
\rjpsi = \frac{\mathcal{B}(\bctaunu)}{\mathcal{B}(\bcmunu)},
\end{equation}
for which the central values of the current SM predictions are in the range of 0.25 to 0.28, where the spread arises from the choice of modeling approach for form factors~\cite{Anisimov:1999jx,Kiselev:2002vz,Ivanov:2006ni,Hernandez:2006gt}.
Here and throughout the Letter charge-conjugate processes are implied.

The measurement is performed using data recorded with the LHCb detector at the Large Hadron Collider in 2011 and 2012,  corresponding to integrated luminosities of 1\invfb and 2\invfb collected at proton-proton ($pp$) center-of-mass energies of $7\tev$ and $8\tev$, respectively. 
The analysis procedure is designed to identify both the signal decay chain \bctaunu\ and the normalization mode \bcmunu, with $\jpsi \to \mup \mun$ and $\taup\to\mup\neum\neutb$, through their identical visible final states $(\mup \mun)\mup$.
The muon candidate not originating from the \jpsi\ is referred to as the \textit{unpaired} muon.
The two modes are distinguished using differences in their kinematic properties.
The selected sample contains contributions from the signal and the normalization modes, as well as several background processes.
The contributions of the various components are determined from a multidimensional fit to the data, where each component is represented by a template distribution derived from control data samples or from simulation validated against data.
The selection and fit procedures are developed without knowledge of the signal yield (blinded).

The \lhcb detector is a single-arm forward spectrometer covering the pseudorapidity range $2<\eta <5$, described in detail in Refs.~\cite{Alves:2008zz,LHCb-DP-2014-002}.
Notably for this analysis, muons are identified by a system composed of alternating layers of iron and multiwire proportional chambers~\cite{LHCb-DP-2012-002}.
The online event selection is performed by a trigger~\cite{LHCb-DP-2012-004}, which in this case consists of a hardware stage, based on information from the calorimeter and muon systems, followed by a software stage, which applies a full event reconstruction.
Simulated data samples, which are used for producing fit templates and evaluating the signal to normalization efficiency ratio, are produced using the software described in Refs.~\cite{Sjostrand:2006za,*Sjostrand:2007gs,LHCb-PROC-2010-056,Chang:2005hq,Allison:2006ve, *Agostinelli:2002hh}.

Events containing a \jpsimu\ candidate are required to have been selected by the LHCb hardware dimuon trigger.
In the software trigger, the events are required to meet criteria designed to select $\jpsi\to\mun\mup$ candidates constructed from oppositely charged tracks whose particle identification information is consistent with a muon.
The \jpsi candidate must have $\pt> 2\gevc$, where \pt is the component of the momentum transverse to the beam, and have a reconstructed mass consistent with the known \jpsi mass \cite{PDG2017}.
In addition, the momenta of the \jpsi decay products must each exceed 5\gevc and at least one muon candidate must have $\pt> 1.5 \gevc$.
In the offline reconstruction, the decay products of the \jpsi candidate must match the the muon candidates responsible for the trigger.
 
Further requirements are imposed in the offline selection, including ones imposed to ensure good-quality tracks.
The $\jpsi$ candidate is required to have well-identified muon decay products, to have a decay vertex significantly separated from any primary vertex in the event, and to have an invariant mass within 55\mevcc of the known \jpsi mass.
A veto is applied to exclude candidates in which the invariant mass of the opposite-sign muon pair formed by swapping the unpaired muon with a muon from the \jpsi candidate is consistent with the \jpsi mass.
The unpaired muon candidate must have $\pt>750\mevc$ and be significantly separated from any PV.
It is required to satisfy muon identification criteria, have a momentum in the range  $3 <p< 100 \gevc$, and be in the pseudorapidity range 2 to 5.
The \jpsi\ candidate and unpaired muon are required to form a vertex with the $\jpsi$ candidate, using loose criteria to reduce any inefficiency due to the \jpsi--\mup separation induced by the \taup flight in the signal decay.
To suppress combinatorial background constructed from the decay products of the other \bquark\ hadrons in the event, the \jpsi\ and the unpaired $\mu$ candidates must not have momenta pointing in nearly opposite directions in the plane transverse to the beam axis.
In the rare ($<2\%$) events where more than one candidate is selected, a single candidate is retained randomly but reproducibly. 

The \jpsimu\ candidates from partially reconstructed \bquark-hadron decays, including \Bc decays to a $\jpsi H_c$ pair, where $H_c$ stands for a charmed hadron,  and semileptonic $\Bc\,\to\,\jpsi (\textrm{n}\pi)\mup\neum$ decays with $\textrm{n} \geq 2$, are typically accompanied by additional nearby charged particles.
 In order to suppress these background contributions, candidates are required to be isolated from additional tracks in the event based on a boosted decision tree (BDT) described in Ref.~\cite{LHCb-PAPER-2015-025}.
The algorithm assigns a score based on whether a given track is likely to have originated from the signal \Bc candidate or from the rest of the event.
The signal sample is constructed by requiring that no tracks in the event are consistent with originating from the \Bcp\ candidate, based on their BDT response value, and is thus enriched in \bctaunu and \bcmunu decays.

The selection efficiencies for the signal and normalization modes are determined from simulation.
To account for the effect of differing detector occupancy and resolution between simulation and data, the joint distributions of the track multiplicity and the significances of the separation of
the \jpsi and of the unpaired muon from the associated PV (defined to be the PV with respect to which the particle under consideration has the smallest impact parameter $\chi^2$, which is the difference in $\chi^2$ of the PV fit with and without the particle in question) in the simulated samples are weighted to match the observed distribution in a subset of the data sample enriched in the normalization mode, without biasing the distribution of the simulated decay time (\ie, the proper time elapsed between the production and decay of the \Bcp\ meson)~\cite{Rogozhnikov:2016bdp}.
This subset is created by excluding events with positive missing mass or long decay times and increasing the rejection of partially reconstructed events using the isolation BDT.
The ratio of the signal efficiency to that of the normalization mode in the nominal selection is found to be  $(52.4\pm 0.4)\%$, where the uncertainty reflects the limited size of the simulation samples.

The differences in the kinematic distributions of the various processes are exploited to disentangle their respective contributions to the selected \jpsimu\ sample.
The large $\mu$--$\tau$ mass difference and the presence of extra neutrinos from the decay $\taup\to\mup\neum\neutb$ result in distinct distributions for the signal relative to the normalization mode. 
Three kinematic quantities are used: the unpaired-muon energy in the $\Bc$ rest frame, $E_{\mu}^*$; the missing mass squared, defined as
 $\mmsq=(p_{\Bc} \!-\! p_{\jpsi} \!-\! p_{\mu})^2$; and the squared four-momentum transfer to the lepton system, $q^2=(p_{\Bc} - p_{\jpsi})^2$, where $p_{\Bc}$, $p_{\jpsi}$ and $p_{\mu}$ are the four-momenta of the \Bc meson, the $\jpsi$ meson, and the unpaired muon, respectively.
These quantities are approximated using a technique developed in Ref.~\cite{LHCb-PAPER-2015-025} that estimates the \Bcp\ momentum despite the presence of one or more missing neutrinos, using the flight direction of the candidate, determined from the vector joining the associated PV and the decay vertex, and the momenta of its decay products.
The lifetime of the \Bcp meson, which is nearly three times shorter than that of other \bquark hadrons, provides an additional handle for discriminating against the large background that originates from lighter \bquark hadrons. 
The decay time for each $\jpsi\mup$ candidate is approximated using the decay distance of the candidate, determined from the approximated \Bcp\ momentum vector and the displacement of its reconstructed vertex relative to its associated PV.
 
The contributions of various components to the sample of \jpsimu\ candidates are represented by three-dimensional histogram templates, binned in \mmsq, the decay time of the \Bcp\ candidate, and a categorical quantity $Z$, representing eight bins in $(\emu,q^2)$.
The values $0$--$3$ of $Z$ correspond to bins where $q^2 < \SI{7.15}{\gevgevcccc}$ and \emu\ is divided with thresholds at $[0.68,1.15,1.64] \, \si{\giga \electronvolt}$.
The values $4$--$7$ correspond to bins with the same \emu\ ranges, but where $q^2 \geq \SI{7.15}{\gevgevcccc}$.
These multidimensional histograms reflect nontrivial correlations among the three quantities.
The sources of the components represented in the fit, and the procedures used to obtain their corresponding templates from simulation and data, are outlined below.

The templates are derived from simulation for the signal and the normalization modes, which requires knowledge of the $\Bc\,\to\,\jpsi \ell^+ \nu_\ell$ form factors. 
These have not yet been precisely determined and the theoretical predictions, \eg\ those from Refs.~\cite{Kiselev:2002vz} and \cite{Ebert:2003cz}, are yet to be tested against data.
Thus, for this measurement, the shared form factors for the signal and normalization modes are determined directly from the data by employing a $z$-expansion parametrization inspired by Ref.~\cite{Bourrely:2008dg} to fit a subsample of the data excluding events with missing mass greater than $1\gevgevcccc$.
In this expansion, the form factors $V(q^2)$, $A_0(q^2)$, $A_1(q^2)$, and $A_2(q^2)$ (following the convention of Ref.~\cite{Ebert:2003cz}) are fit by functions of the form
\begin{equation}
f(q^2) = \frac{1}{1-q^2/{M_\mathrm{pole}^2}} \sum_{k=0}^K a_k z(q^2)^k,
\end{equation}
where $z(q^2)$ is defined in Ref.~\cite{Bourrely:2008dg}.
The pole mass $M_\mathrm{pole}$ is the mass of the excited \Bcp\ state with quantum numbers corresponding to the form factor: the $J^P = 1^-$ state for the form factor $V(q^2)$, taken to be \SI{6.33}{\gevcc}; the $0^-$ state for $A_0(q^2)$, which is the \Bcp\ mass itself; and finally the $1^+$ state for $A_1(q^2)$ and $A_2(q^2)$, taken to be \SI{6.73}{\gevc} \cite{Ebert:2003cz,Kiselev:2002vz}.
The form factor $A_2(q^2)$ is fit to $K=0$ order, while the others are fit to the linear $K=1$ order.
The parameters $a_k$ obtained from this procedure contain the effects of the reconstruction resolution of the kinematic parameters and cannot be directly compared with existing theoretical predictions.

Simulation is used to determine the templates for the feed-down processes $\Bcp\,\to\,\psitwos \mup \neum$,  $\Bcp\,\to\,\psitwos \taup \neut$, $\Bcp\,\to\,\chicone \mup \neum$, and $\Bcp\,\to\,\chictwo \mup \neum$, and backgrounds from $\Bcp\,\to\,\jpsi H_c X$.
The last is represented by a cocktail of decays that result from $b\,\to\,c \bar{c} s$  transitions.
The branching fractions for the decays $\jpsi\,\to\,\mup\mun$, $\psitwos\,\to\,\jpsi X$, $\chi_{c(1,2)}\,\to\,\jpsi \gamma$, and $\taup\,\to\,\mup \neum \neutb$ are fixed to the known values \cite{PDG2017}.
A possible feed-down contribution from $\Bcp \to X(3872)\mup \neum$, where the $X(3872)$ state decay produces a $\jpsi$, is considered in the determination of the systematic uncertainties.
The semimuonic  \Bcp decays to \chicone\ and \chictwo\ modes are constrained to have the same branching fractions relative to the normalization mode, differing only due to the respective branching fractions of $\chi_{c(1,2)}$ to $\jpsi \gamma$, consistent with theoretical expectations \cite{PhysRevD.79.114018}.  
Form factors for these decays are taken from Ref.~\cite{PhysRevD.79.114018}.
The rare decay $\Bcp\,\to\,\chiczero\ \mup \neum$ (suppressed by the low $\chiczero\,\to\,\jpsi X$ branching fraction) and semitauonic decays involving $\chi_c$ states are neglected and are accounted for in the systematic uncertainties.

The background processes $\Bcp\,\to\,\jpsi H_c X$ are modeled using a cocktail of two-body decays and quasi-two-body decays that proceed through excited $\Ds$ resonances.
Several decay modes in the cocktail have recently been measured at LHCb \cite{LHCb-PAPER-2013-010}, and for others the branching fractions are fixed by analogy to the well measured $B\,\to\,D^* H_c X$ decays \cite{PDG2017}. The cocktail consists of the two-body and quasi-two-body decays in equal proportion.

The decay-time distributions derived from simulated \Bcp\ decays are corrected for acceptance differences between data and simulation.
This is achieved by applying weights to the simulated distribution from a study of a control sample of $\jpsi \Kp$ combinations from the decay $B^0\to\jpsi K^{*}(892)^0$, with $K^{*}(892)^0\to \Kp\pim$, in data and simulation; the weights are calculated in bins of the decay time and of the relative momentum carried by the $\pim$ omitted from the combination (analogous to that of the unobserved neutrino(s) in the simulation samples).
The \Bcp\ lifetime is allowed to vary in the fit, constrained by its measured value and precision.

The combinatorial background in the selected \jpsimu\ sample is predominantly due to \jpsi\ mesons from $B_{u,d,s}\,\to\,\jpsi X$ decays paired with muon candidates from the rest of the event.
This background source is modeled using a set of three template histograms taken from simulation for the three $B$-meson species, with their relative fractions constrained in accordance with the production cross-sections and their respective branching fractions.
A fit is performed to the \jpsimu\ mass distribution above $6.4\gevcc$, higher than the $\Bc$ mass, to validate the modeling of this background and correct for possible sources of combinatorial background in data unaccounted for by the model, including decays of $\bquark$ baryons and the effect of unknown branching fractions.
A linear correction to the \jpsimu\ mass distribution in the simulation is determined by this fit and applied to the combinatorial background templates, and is varied within bounds to determine a systematic uncertainty.

A separate background comes from pairing unrelated muons to form \jpsi\ candidates.
The template for this combinatorial \jpsi\ component is determined using events where the \jpsi\ invariant mass lies above the nominal selection threshold, with its normalization fixed using a fit to the $\mup\mun$ invariant mass distribution. 
Two models for the shape of the combinatorial background in the \jpsi mass distribution are considered. 
The nominal fit uses a mixture of distributions with Gaussian cores and power law tails \cite{Skwarnicki:1986xj} for the true $\jpsi\,\to\,\mup \mun$ component and an exponential function for the combinatorial background. 
An alternative fit is performed to evaluate a corresponding systematic uncertainty.
 
The largest background component is due to the inclusive decays of light \bquark hadrons to $\jpsi$ mesons, in which an accompanying pion or kaon (or, less frequently, proton or electron) is misidentified as a muon, hereafter referred to as the mis-ID background.
A data-driven approach is used to construct templates for this background component.  
A sample of $\jpsi h^+$ candidates, where $h^+$ stands for a charged hadron, is selected following similar criteria to those of the signal sample, but with the $h^+$ failing the muon identification criteria.
This control sample is enriched in various hadron species (primarily pions, kaons, and protons) and electrons.
Using several high-purity control samples of identified hadrons, weights are computed that represent the probability that a hadron with particular kinematic properties would pass the muon criteria.
These weights are applied to the $\jpsi h^+$ sample to generate binned templates representing these background components.
The normalization of each of these components is allowed to vary in the fit to the data.

A binned maximum likelihood (ML) fit is performed using the templates representing the various components.
The number of candidates from each component, with the exception of the combinatorial \jpsi\ background, are allowed to vary in the fit, as are the shape parameters corresponding to the \Bcp\ lifetime and the $A_0(q^2)$ form factor.
The contributions of the feed-down processes involving the decays of higher-mass charmonium states, $\Bcp\,\to\,\psitwos \mup \neum$,  
$\Bcp\,\to\,\chi_{c(0,1,2)}(1P) \mup \neum$ are allowed to vary in the fit, whereas the ratio of the branching fractions $\mathcal{R}(\psitwos)=\mathcal{B}(\Bc \to\psitwos\taup\neut)/\mathcal{B}(\Bc \to\psitwos\mup\neum)$ is fixed to the predicted SM value of $8.5\%$ \cite{Kiselev:2002vz}.
This is later varied for the evaluation of a systematic uncertainty.

Extensive studies of the fit procedure are carried out to identify potential sources of bias in the fit. 
Simulated signal is added to the data histograms, and the resulting changes in the value of \rjpsi\ from the fit are found to be consistent with the injected signal increments.
The procedure is also applied to the mis-ID background, which shows no bias in the fitted number of events as a function of injected events. 
Another important consideration for this measurement is the disparate properties of the various templates.
Some templates are populated in all kinematically allowed bins, such as the mis-ID background that is derived from large data samples.
Others are sparsely populated and contain empty bins, \eg\ for modes with low efficiency and yields that are obtained from simulated events.
Pseudoexperiments with template compositions similar to those in this analysis reveal a possible bias of the fit results.
Hence, the binning scheme for this analysis is chosen to minimize the number of empty bins in the sparsely populated templates, while retaining the discriminating power of the distributions.
Kernel density estimation (KDE)~\cite{Cranmer:2000du} is used to derive continuous distributions representative of the nominal fit templates.
Simulated pseudoexperiments using histogram templates sampled from these continuous distributions are then used to evaluate any remaining bias that results.
Based on these studies a Bayesian procedure is implemented for correcting the raw \rjpsi\ value  after unblinding.

\begin{figure}[tb]
\begin{center}
\ifthenelse{\boolean{usetwocolumn}}{
\includegraphics[width=0.8\linewidth]{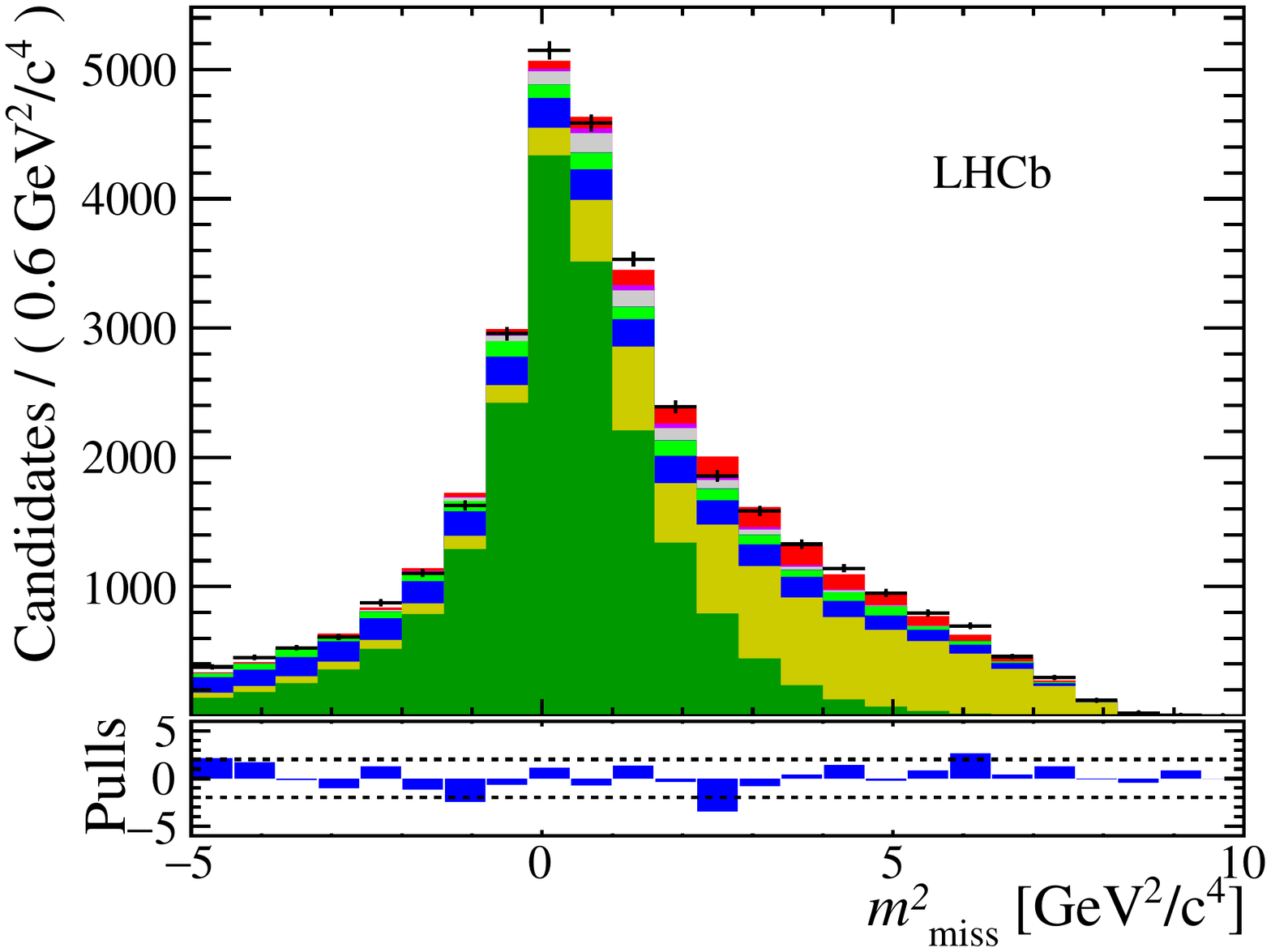}

\includegraphics[width=0.8\linewidth]{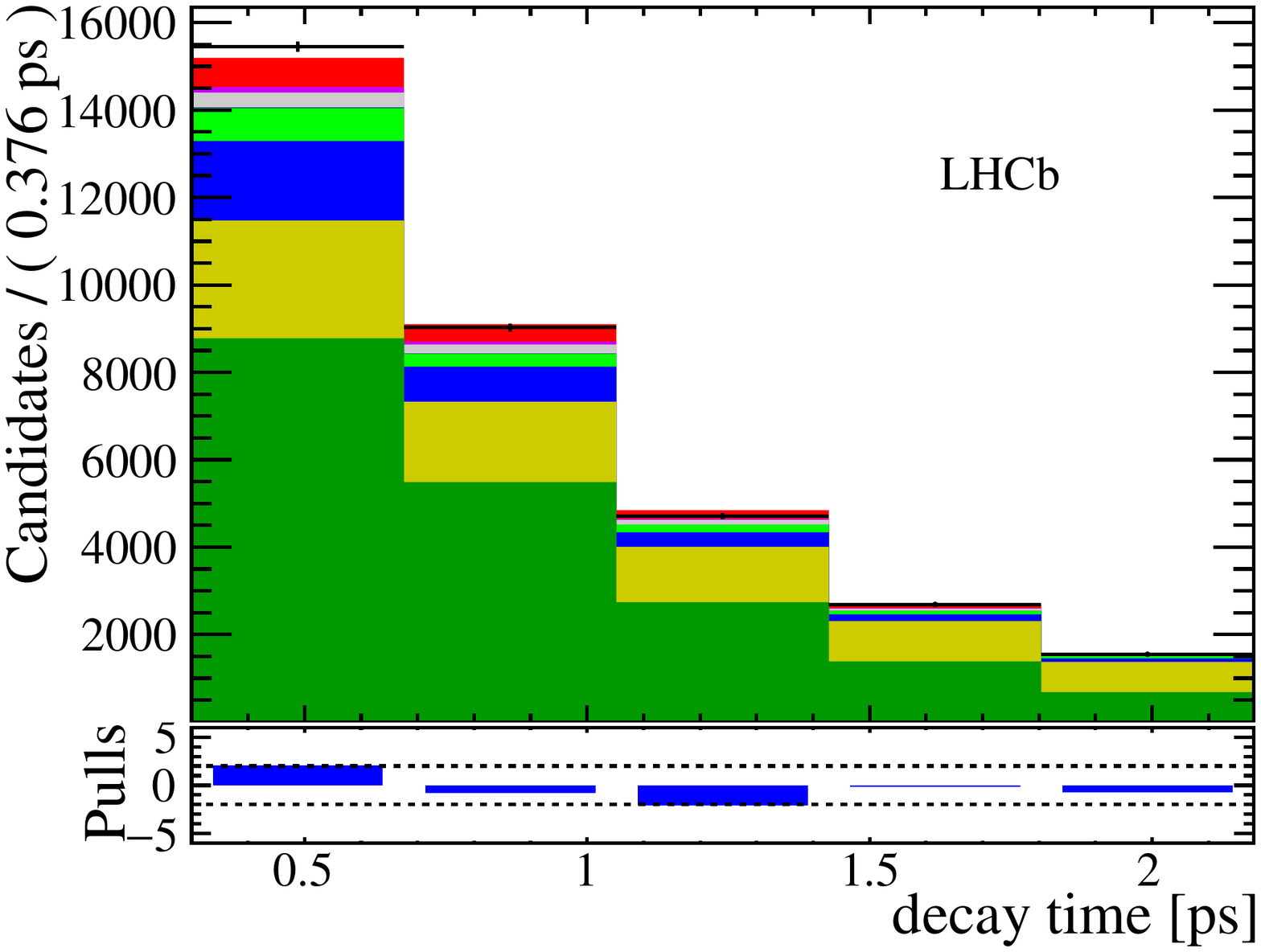}

\includegraphics[width=0.8\linewidth]{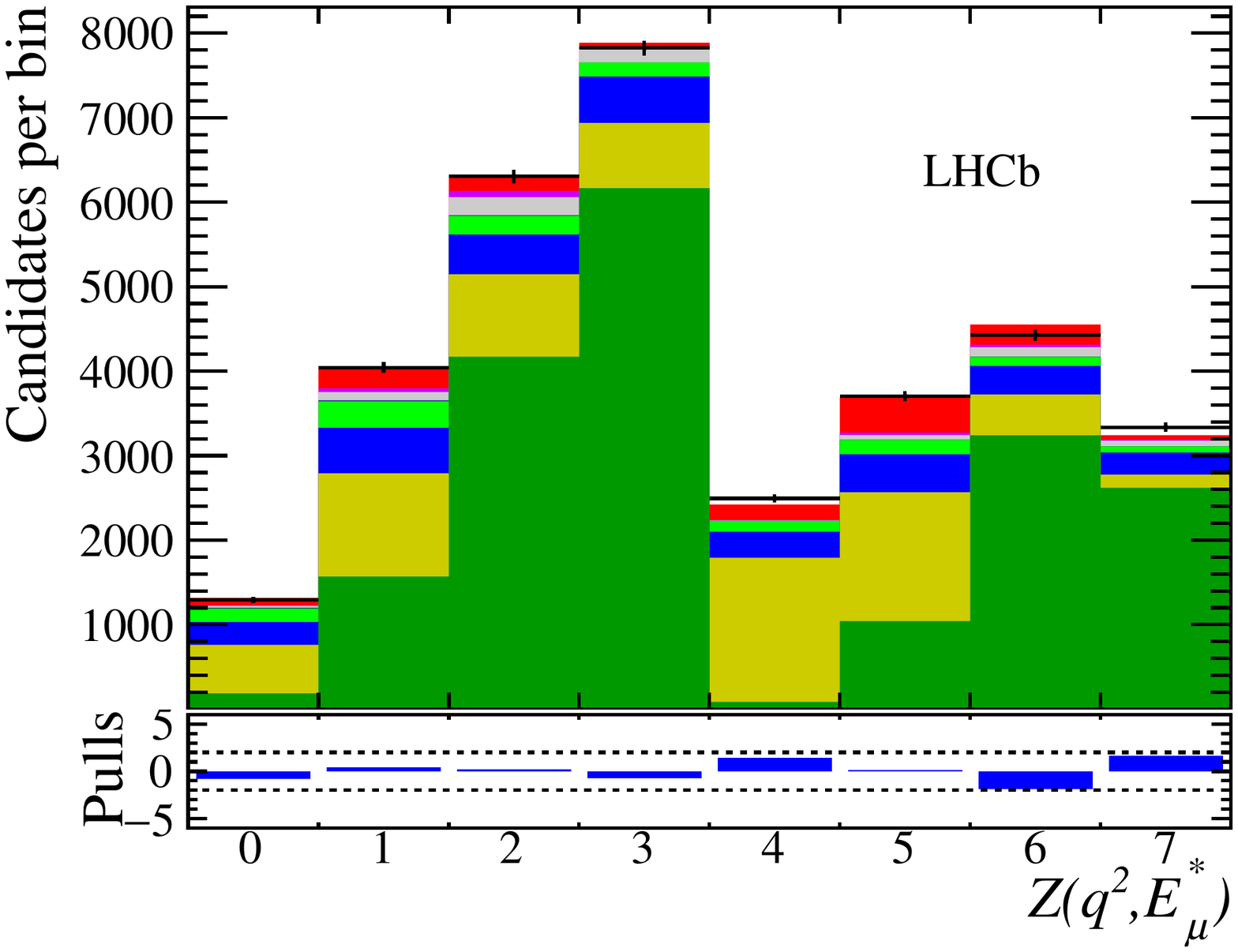}

\includegraphics[width=0.8\linewidth]{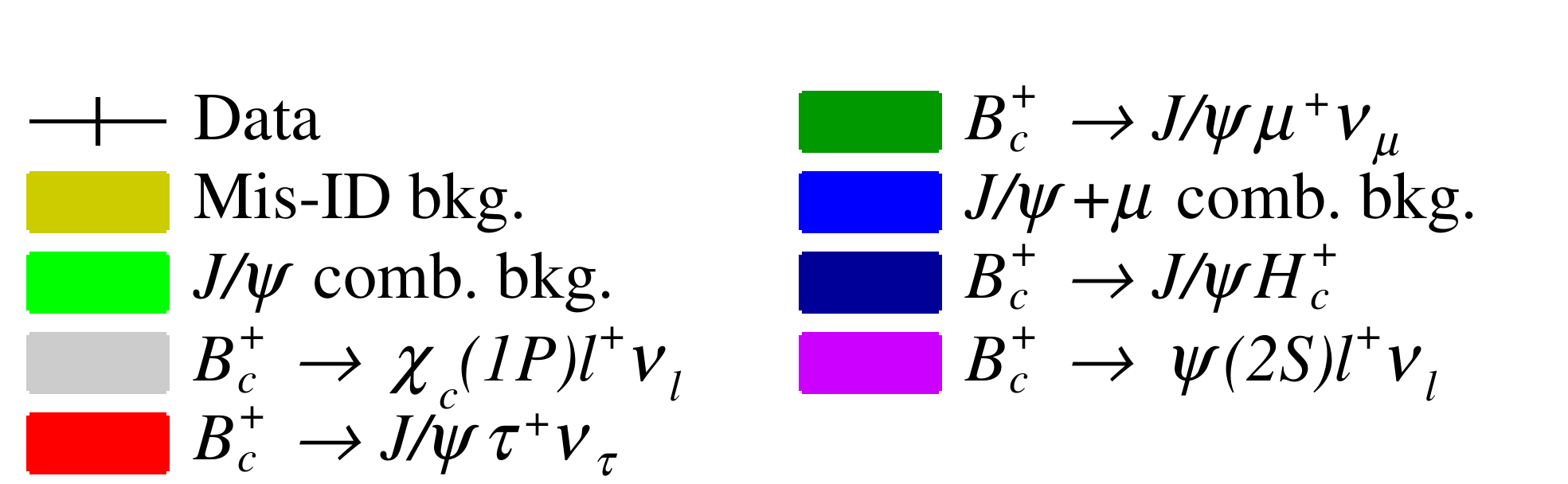}
}{
\includegraphics[width=0.45\linewidth]{Fig1a.pdf}

\includegraphics[width=0.45\linewidth]{Fig1b.pdf}

\includegraphics[width=0.45\linewidth]{Fig1c.pdf}

\includegraphics[width=0.45\linewidth]{Fig1_legend.pdf}
}
\end{center}
\caption{Distributions of (top) \mmsq, (middle) decay time, and (bottom) $Z$ of the signal data, overlaid with projections of the fit model with all normalization and shape parameters at their best-fit values. Below each panel differences between the data and fit are shown, normalized by the Poisson uncertainty in the data; the dashed lines are at the values $\pm 2$.}
\label{fig:theFit}
\end{figure}

The results of the fit are presented in Fig. \ref{fig:theFit}, showing the projections of the nominal fit result onto the quantities \mmsq, decay time, and $Z$.
The fit yields $1400\pm300$ signal and $19140\pm340$ normalization decays, where the errors are statistical and correlated.
Accounting for the $\taup\to\mup\neum\neutb$ branching fraction and the ratio of efficiencies ($(52.4\pm 0.4)\%$) gives an uncorrected value of $0.79$ for \rjpsi.
Correcting for the mean expected bias at this value, we obtain $\rjpsi=0.71 \pm 0.17 \stat$.
The significance of the signal, determined from a likelihood scan procedure and corrected for the systematic uncertainty, is found to be 3 standard deviations.

\begin{table}
\caption[Systematic uncertainties on \rjpsi]{Systematic uncertainties in the determination of \rjpsi.}
\label{table:systematics}
\setlength\tabcolsep{0pt}
\centering
\begin{tabular*}{\columnwidth}{@{\extracolsep{\fill}}lr}
  \hline
  \textbf{Source of uncertainty} & \textbf{Size ($\mathbf{\times 10^{-2}}$)} \\
  \hline
  \ifthenelse{\boolean{usetwocolumn}}{Finite simulation size}{Limited size of simulation samples} & 8.0 \\ %
  $\Bcp\,\to\,\jpsi$ form factors & 12.1 \\ %
  $\Bcp\,\to\,\psitwos$ form factors & 3.2 \\ %
  Fit bias correction & 5.4 \\
  $Z$ binning strategy & 5.6 \\
  \ifthenelse{\boolean{usetwocolumn}}{Misid. bkg. strategy}{Misidentification background strategy} & 5.6 \\ %
  \ifthenelse{\boolean{usetwocolumn}}{Comb. bkg. cocktail}{Combinatorial background cocktail} & 4.5 \\
  \ifthenelse{\boolean{usetwocolumn}}{Comb. \jpsi\ bkg. scaling}{Combinatorial \jpsi\ sideband scaling} & 0.9 \\
   $\Bcp\,\to\,\jpsi H_c X $ contribution & 3.6 \\  %
  \ifthenelse{\boolean{usetwocolumn}}{\psitwos\ and \chic\ feed-down}{Semitauonic \psitwos\ and \chic\ feed-down} & 0.9 \\ %
  Weighting of simulation samples & 1.6 \\
  Efficiency ratio  & 0.6 \\ %
  $\mathcal{B}(\taup\,\to\,\mup \neum \neutb)$ & 0.2 \\ %
  \ifthenelse{\boolean{usetwocolumn}}{\textbf{Systematic uncertainty}}{\textbf{Total systematic uncertainty}} & \textbf{17.7} \\
  \hline
  \textbf{Statistical uncertainty} & \textbf{17.3} \\
  \hline
\end{tabular*}
\end{table}

Systematic uncertainties on \rjpsi\ are listed in Table  \ref{table:systematics}.
The effect of the limited size of the simulated samples on the template shapes is determined using the procedure of Refs.~\cite{Barlow:1993dm,CERN-OPEN-2012-016}.
In the nominal fit, the $\Bcp\,\to\,\jpsi$ form factor parameters, except for the scalar form factor that primarily affects the semitauonic mode, are fixed to the values obtained from a fit to a subset of the data enriched in the normalization mode.
To assess the effect on \rjpsi\ due to this procedure, an alternative fit is performed with the form factor parameters allowed to vary, and the difference in quadrature of the uncertainties is assigned as a systematic uncertainty.
The effect due to the $\Bcp\,\to\,\psitwos$ form factors is evaluated by comparing fits using two different theoretical models for this template~\cite{Kiselev:2002vz,Ebert:2003cz}.

The systematic uncertainty of the bias correction is calculated from the difference in bias between fits to simulated data based on a set of realistic parametrized distributions and corresponding fits based on KDE versions of these distributions.  
The effect of the placement of the bin thresholds in the quantity $Z$ is determined by varying the boundaries of the thresholds in \emu\ and $q^2$, and by reducing the number of bins in the fit.
The data-driven method employed to determine the mis-ID background is repeated with an alternate approach for modeling the effect of misreconstructed tracks within the mis-ID control sample (rejected from the nominal sample by muon PID requirements).
The fit procedure is repeated with templates derived from this alternate method, and an uncertainty is assigned using half the difference between the resulting central value of \rjpsi\ and the nominal value.
The systematic uncertainty due to the combinatorial background model is determined by varying the linear correction made to its \jpsimu\ mass distribution, described above, within its bounds.
The uncertainty due to the combinatorial background in the $\jpsi$ peak region is determined by varying the normalization of this component within the range determined from the alternative fit to the invariant-mass distribution of \jpsi\ candidates. 

The systematic uncertainty due to the contribution of the process $\Bcp\,\to\,\jpsi H_c X$, which is poorly resolved by the fit, is determined by fixing the yield relative to the normalization to that expected from the estimated branching fraction for these decays~\cite{PDG2017, LHCb-PAPER-2013-010}.
The effect of fixing the contribution of the semitauonic decay $\Bcp\,\to\,\psitwos \taup \neum$ is determined by varying $\mathcal{R}(\psitwos)$  by $\pm$50\% of the predicted value.
Background from the feed-down decays $\Bcp\,\to\,X(3872)\mup \neum$, with the principal decay chains $X(3872)\to \jpsi\pip\pim$ and $X(3872)\to \jpsi\gamma$, is kinematically similar to the background from $\Bcp\,\to\,\psitwos \taup \neum$.
An approximate bound on the number of $X(3872)$ candidates in the sample is obtained from the invariant mass distribution of $\jpsi\pip\pim$ combinations in the sample.
This bound is found to be less than the uncertainty in the \psitwos\ yield, and thus no additional uncertainty is assigned.
In general, the effect of charmonium states above the open-charm threshold, which have large total width, are negligible as a result of their small decay rate to final states containing $\jpsi$.
The uncertainty due to the small contribution of semitauonic decays involving \chic\ states is assessed by assuming that the entire yield for this mode is absorbed in the signal mode, and is summed in quadrature with that from the \psitwos\ feed-down mode.

The systematic uncertainty due to the weighting of the simulation distributions of event parameters (the track multiplicity and the separation significances of the \jpsi and of the unpaired muon) is determined by varying the criteria for the definition of the subset of the data sample enriched in the normalization mode used in the weighting procedure, and employing alternative methods to account for the misidentified muon candidates in the sample.
The uncertainty in the efficiency ratio measured in simulation is propagated to \rjpsi, and is dominated by the statistical uncertainty of the simulation sample.

In summary,  the decay $\Bcp\,\to\,\jpsi \tau^+ \neut$ is studied using data corresponding to $3 \invfb$ recorded with the LHCb detector during 2011 and 2012, leading 
to the first measurement of the ratio of branching fractions
\ifthenelse{\boolean{usetwocolumn}}{
\begin{equation}\begin{split}
\rjpsi &= \frac{\mathcal{B}(\Bcp\,\to\,\jpsi \tau^+ \neut)}{\mathcal{B}(\Bcp\,\to\,\jpsi \mup \neum)} \\
&= 0.71 \pm 0.17 \stat\,  \pm 0.18\syst.\end{split}\end{equation}
}{
    \begin{equation}\rjpsi = \frac{\mathcal{B}(\Bcp\,\to\,\jpsi \tau^+ \neut)}{\mathcal{B}(\Bcp\,\to\,\jpsi \mup \neum)} = 0.71 \pm 0.17 \stat\,  \pm 0.18\syst.\end{equation}
}
This result lies within 2 standard deviations of the range of central values currently predicted from the Standard Model, 0.25 to 0.28.

\section*{Acknowledgements}

\noindent We express our gratitude to our colleagues in the CERN
accelerator departments for the excellent performance of the LHC. We
thank the technical and administrative staff at the LHCb
institutes. We acknowledge support from CERN and from the national
agencies: CAPES, CNPq, FAPERJ and FINEP (Brazil); MOST and NSFC
(China); CNRS/IN2P3 (France); BMBF, DFG and MPG (Germany); INFN
(Italy); NWO (The Netherlands); MNiSW and NCN (Poland); MEN/IFA
(Romania); MinES and FASO (Russia); MinECo (Spain); SNSF and SER
(Switzerland); NASU (Ukraine); STFC (United Kingdom); NSF (USA).  We
acknowledge the computing resources that are provided by CERN, IN2P3
(France), KIT and DESY (Germany), INFN (Italy), SURF (The
Netherlands), PIC (Spain), GridPP (United Kingdom), RRCKI and Yandex
LLC (Russia), CSCS (Switzerland), IFIN-HH (Romania), CBPF (Brazil),
PL-GRID (Poland) and OSC (USA). We are indebted to the communities
behind the multiple open-source software packages on which we depend.
Individual groups or members have received support from AvH Foundation
(Germany), EPLANET, Marie Sk\l{}odowska-Curie Actions and ERC
(European Union), ANR, Labex P2IO, ENIGMASS and OCEVU, and R\'{e}gion
Auvergne-Rh\^{o}ne-Alpes (France), RFBR and Yandex LLC (Russia), GVA,
XuntaGal and GENCAT (Spain), Herchel Smith Fund, the Royal Society,
the English-Speaking Union and the Leverhulme Trust (United Kingdom).

\addcontentsline{toc}{section}{References}
\setboolean{inbibliography}{true}
\bibliographystyle{LHCb}
\bibliography{main,LHCb-PAPER,LHCb-CONF,LHCb-DP,LHCb-TDR}

\ifx\mcitethebibliography\mciteundefinedmacro
\PackageError{LHCb.bst}{mciteplus.sty has not been loaded}
{This bibstyle requires the use of the mciteplus package.}\fi
\providecommand{\href}[2]{#2}
\begin{mcitethebibliography}{10}
\mciteSetBstSublistMode{n}
\mciteSetBstMaxWidthForm{subitem}{\alph{mcitesubitemcount})}
\mciteSetBstSublistLabelBeginEnd{\mcitemaxwidthsubitemform\space}
{\relax}{\relax}

\bibitem{Lees:2012xj}
BaBar collaboration, J.~P. Lees {\em et~al.},
  \ifthenelse{\boolean{articletitles}}{\emph{{Evidence for an excess of $\Bbar
  \to D^{(*)} \tau^-\neutb$ decays}},
  }{}\href{http://dx.doi.org/10.1103/PhysRevLett.109.101802}{Phys.\ Rev.\
  Lett.\  \textbf{109} (2012) 101802},
  \href{http://arxiv.org/abs/1205.5442}{{\normalfont\ttfamily
  arXiv:1205.5442}}\relax
\mciteBstWouldAddEndPuncttrue
\mciteSetBstMidEndSepPunct{\mcitedefaultmidpunct}
{\mcitedefaultendpunct}{\mcitedefaultseppunct}\relax
\EndOfBibitem
\bibitem{Lees:2013uzd}
BaBar collaboration, J.~P. Lees {\em et~al.},
  \ifthenelse{\boolean{articletitles}}{\emph{{Measurement of an excess of
  $\Bbar \to D^{(*)}\tau^- \neutb$ decays and implications for charged Higgs
  bosons}}, }{}\href{http://dx.doi.org/10.1103/PhysRevD.88.072012}{Phys.\ Rev.\
   \textbf{D88} (2013) 072012},
  \href{http://arxiv.org/abs/1303.0571}{{\normalfont\ttfamily
  arXiv:1303.0571}}\relax
\mciteBstWouldAddEndPuncttrue
\mciteSetBstMidEndSepPunct{\mcitedefaultmidpunct}
{\mcitedefaultendpunct}{\mcitedefaultseppunct}\relax
\EndOfBibitem
\bibitem{Huschle:2015rga}
Belle collaboration, M.~Huschle {\em et~al.},
  \ifthenelse{\boolean{articletitles}}{\emph{{Measurement of the branching
  ratio of $\Bbar \to D^{(\ast)} \tau^- \neutb$ relative to $\Bbar \to
  D^{(\ast)} \ell^- \neulb$ decays with hadronic tagging at Belle}},
  }{}\href{http://dx.doi.org/10.1103/PhysRevD.92.072014}{Phys.\ Rev.\
  \textbf{D92} (2015) 072014},
  \href{http://arxiv.org/abs/1507.03233}{{\normalfont\ttfamily
  arXiv:1507.03233}}\relax
\mciteBstWouldAddEndPuncttrue
\mciteSetBstMidEndSepPunct{\mcitedefaultmidpunct}
{\mcitedefaultendpunct}{\mcitedefaultseppunct}\relax
\EndOfBibitem
\bibitem{Sato:2016svk}
Belle collaboration, Y.~Sato {\em et~al.},
  \ifthenelse{\boolean{articletitles}}{\emph{{Measurement of the branching
  ratio of $\Bzb \rightarrow D^{*+} \tau^- \neutb$ relative to $\Bzb
  \rightarrow D^{*+} \ell^- \neulb$ decays with a semileptonic tagging
  method}}, }{}\href{http://dx.doi.org/10.1103/PhysRevD.94.072007}{Phys.\ Rev.\
   \textbf{D94} (2016) 072007},
  \href{http://arxiv.org/abs/1607.07923}{{\normalfont\ttfamily
  arXiv:1607.07923}}\relax
\mciteBstWouldAddEndPuncttrue
\mciteSetBstMidEndSepPunct{\mcitedefaultmidpunct}
{\mcitedefaultendpunct}{\mcitedefaultseppunct}\relax
\EndOfBibitem
\bibitem{Hirose:2016wfn}
Belle collaboration, S.~Hirose {\em et~al.},
  \ifthenelse{\boolean{articletitles}}{\emph{{Measurement of the $\tau$ lepton
  polarization and $R(D^*)$ in the decay $\Bbar \to D^* \tau^- \neutb$}},
  }{}\href{http://dx.doi.org/10.1103/PhysRevLett.118.211801}{Phys.\ Rev.\
  Lett.\  \textbf{118} (2017) 211801},
  \href{http://arxiv.org/abs/1612.00529}{{\normalfont\ttfamily
  arXiv:1612.00529}}\relax
\mciteBstWouldAddEndPuncttrue
\mciteSetBstMidEndSepPunct{\mcitedefaultmidpunct}
{\mcitedefaultendpunct}{\mcitedefaultseppunct}\relax
\EndOfBibitem
\bibitem{Hirose:2017dxl}
Belle collaboration, S.~Hirose {\em et~al.},
  \ifthenelse{\boolean{articletitles}}{\emph{{Measurement of the $\tau$ lepton
  polarization and $R(D^*)$ in the decay $\Bbar \rightarrow D^* \tau^- \neutb$
  with one-prong hadronic $\tau$ decays at Belle}},
  }{}\href{http://arxiv.org/abs/1709.00129}{{\normalfont\ttfamily
  arXiv:1709.00129}}\relax
\mciteBstWouldAddEndPuncttrue
\mciteSetBstMidEndSepPunct{\mcitedefaultmidpunct}
{\mcitedefaultendpunct}{\mcitedefaultseppunct}\relax
\EndOfBibitem
\bibitem{LHCb-PAPER-2015-025}
LHCb collaboration, R.~Aaij {\em et~al.},
  \ifthenelse{\boolean{articletitles}}{\emph{{Measurement of the ratio of
  branching fractions
  $\BF(\Bzb\to\Dstarp\taum\neutb)/\BF(\Bzb\to\Dstarp\mun\neumb)$}},
  }{}\href{http://dx.doi.org/10.1103/PhysRevLett.115.111803}{Phys.\ Rev.\
  Lett.\  \textbf{115} (2015) 111803},
  \href{http://arxiv.org/abs/1506.08614}{{\normalfont\ttfamily
  arXiv:1506.08614}}\relax
\mciteBstWouldAddEndPuncttrue
\mciteSetBstMidEndSepPunct{\mcitedefaultmidpunct}
{\mcitedefaultendpunct}{\mcitedefaultseppunct}\relax
\EndOfBibitem
\bibitem{LHCb-PAPER-2017-017}
LHCb collaboration, R.~Aaij {\em et~al.},
  \ifthenelse{\boolean{articletitles}}{\emph{{Measurement of the ratio of the
  $\mathcal{B}(B^0 \to D^{\ast-} \tau^+ \nu_{\tau})$ and
  \hbox{$\mathcal{B}(B^0\to D^{\ast-}\mu^+\nu_{\mu})$} branching fractions
  using three-prong $\tau$-lepton decays}},
  }{}\href{http://arxiv.org/abs/1708.08856}{{\normalfont\ttfamily
  arXiv:1708.08856}}, {submitted to Phys. Rev. Lett}\relax
\mciteBstWouldAddEndPuncttrue
\mciteSetBstMidEndSepPunct{\mcitedefaultmidpunct}
{\mcitedefaultendpunct}{\mcitedefaultseppunct}\relax
\EndOfBibitem
\bibitem{LHCb-PAPER-2017-027}
LHCb collaboration, R.~Aaij {\em et~al.},
  \ifthenelse{\boolean{articletitles}}{\emph{{Measurement of the
  $\mathcal{B}(B^0 \to D^{\ast-} \tau^+ \nu_{\tau})$ branching fraction using
  three-prong $\tau$ decays}},
  }{}\href{http://arxiv.org/abs/1711.02505}{{\normalfont\ttfamily
  arXiv:1711.02505}}, {submitted to Phys. Rev. D}\relax
\mciteBstWouldAddEndPuncttrue
\mciteSetBstMidEndSepPunct{\mcitedefaultmidpunct}
{\mcitedefaultendpunct}{\mcitedefaultseppunct}\relax
\EndOfBibitem
\bibitem{HFLAV}
Heavy Flavor Averaging Group, Y.~Amhis {\em et~al.},
  \ifthenelse{\boolean{articletitles}}{\emph{{Averages of $b$-hadron,
  $c$-hadron, and $\tau$-lepton properties as of summer 2016}},
  }{}\href{http://arxiv.org/abs/1612.07233}{{\normalfont\ttfamily
  arXiv:1612.07233}}, {updated results and plots available at
  \href{http://www.slac.stanford.edu/xorg/hflav/}{{\texttt{http://www.slac.stanford.edu/xorg/hflav/}}}}\relax
\mciteBstWouldAddEndPuncttrue
\mciteSetBstMidEndSepPunct{\mcitedefaultmidpunct}
{\mcitedefaultendpunct}{\mcitedefaultseppunct}\relax
\EndOfBibitem
\bibitem{Tanaka:1994ay}
M.~Tanaka, \ifthenelse{\boolean{articletitles}}{\emph{{Charged Higgs effects on
  exclusive semitauonic $B$ decays}},
  }{}\href{http://dx.doi.org/10.1007/BF01571294}{Z.\ Phys.\ C \textbf{67}
  (1995) 321}, \href{http://arxiv.org/abs/hep-ph/9411405}{{\normalfont\ttfamily
  arXiv:hep-ph/9411405}}\relax
\mciteBstWouldAddEndPuncttrue
\mciteSetBstMidEndSepPunct{\mcitedefaultmidpunct}
{\mcitedefaultendpunct}{\mcitedefaultseppunct}\relax
\EndOfBibitem
\bibitem{Crivellin:2012ye}
A.~Crivellin, C.~Greub, and A.~Kokulu,
  \ifthenelse{\boolean{articletitles}}{\emph{{Explaining $B \to D\tau\nu$, $B
  \to D^{*}\tau\nu$ and $B\to\tau\nu$ in a two Higgs doublet model of type
  III}}, }{}Phys.\ Rev.\  \textbf{D86} (2012) 054014,
  \href{http://arxiv.org/abs/1206.2634}{{\normalfont\ttfamily
  arXiv:1206.2634}}\relax
\mciteBstWouldAddEndPuncttrue
\mciteSetBstMidEndSepPunct{\mcitedefaultmidpunct}
{\mcitedefaultendpunct}{\mcitedefaultseppunct}\relax
\EndOfBibitem
\bibitem{freytsis:2015qca}
M.~Freytsis, Z.~Ligeti, and J.~T. Ruderman,
  \ifthenelse{\boolean{articletitles}}{\emph{{Flavor models for $\Bbar \to
  D^{(*)} \tau \neub$}},
  }{}\href{http://dx.doi.org/10.1103/PhysRevD.92.054018}{Phys.\ Rev.\
  \textbf{D92} (2015) 054018},
  \href{http://arxiv.org/abs/1506.08896}{{\normalfont\ttfamily
  arXiv:1506.08896}}\relax
\mciteBstWouldAddEndPuncttrue
\mciteSetBstMidEndSepPunct{\mcitedefaultmidpunct}
{\mcitedefaultendpunct}{\mcitedefaultseppunct}\relax
\EndOfBibitem
\bibitem{crivellin:2015lwa}
A.~Crivellin, G.~D'Ambrosio, and J.~Heeck,
  \ifthenelse{\boolean{articletitles}}{\emph{{Addressing the LHC flavor
  anomalies with horizontal gauge symmetries}},
  }{}\href{http://dx.doi.org/10.1103/PhysRevD.91.075006}{Phys.\ Rev.\
  \textbf{D91} (2015) 075006},
  \href{http://arxiv.org/abs/1503.03477}{{\normalfont\ttfamily
  arXiv:1503.03477}}\relax
\mciteBstWouldAddEndPuncttrue
\mciteSetBstMidEndSepPunct{\mcitedefaultmidpunct}
{\mcitedefaultendpunct}{\mcitedefaultseppunct}\relax
\EndOfBibitem
\bibitem{LHCb-PAPER-2014-024}
LHCb collaboration, R.~Aaij {\em et~al.},
  \ifthenelse{\boolean{articletitles}}{\emph{{Test of lepton universality using
  $\Bp\to\Kp\ell^+\ell^-$ decays}},
  }{}\href{http://dx.doi.org/10.1103/PhysRevLett.113.151601}{Phys.\ Rev.\
  Lett.\  \textbf{113} (2014) 151601},
  \href{http://arxiv.org/abs/1406.6482}{{\normalfont\ttfamily
  arXiv:1406.6482}}\relax
\mciteBstWouldAddEndPuncttrue
\mciteSetBstMidEndSepPunct{\mcitedefaultmidpunct}
{\mcitedefaultendpunct}{\mcitedefaultseppunct}\relax
\EndOfBibitem
\bibitem{LHCb-PAPER-2017-013}
LHCb collaboration, R.~Aaij {\em et~al.},
  \ifthenelse{\boolean{articletitles}}{\emph{{Test of lepton universality with
  $B^0\to K^{*0} \ell^+ \ell^-$ decays}},
  }{}\href{http://dx.doi.org/10.1007/JHEP08(2017)055}{JHEP \textbf{08} (2017)
  055}, \href{http://arxiv.org/abs/1705.05802}{{\normalfont\ttfamily
  arXiv:1705.05802}}\relax
\mciteBstWouldAddEndPuncttrue
\mciteSetBstMidEndSepPunct{\mcitedefaultmidpunct}
{\mcitedefaultendpunct}{\mcitedefaultseppunct}\relax
\EndOfBibitem
\bibitem{Anisimov:1999jx}
A.~{\relax Yu}. Anisimov, I.~M. Narodetskii, C.~Semay, and B.~Silvestre-Brac,
  \ifthenelse{\boolean{articletitles}}{\emph{{The $B_c$ meson lifetime in the
  light-front constituent quark model}},
  }{}\href{http://dx.doi.org/10.1016/S0370-2693(99)00273-7}{Phys.\ Lett.\
  \textbf{B452} (1999) 129},
  \href{http://arxiv.org/abs/hep-ph/9812514}{{\normalfont\ttfamily
  arXiv:hep-ph/9812514}}\relax
\mciteBstWouldAddEndPuncttrue
\mciteSetBstMidEndSepPunct{\mcitedefaultmidpunct}
{\mcitedefaultendpunct}{\mcitedefaultseppunct}\relax
\EndOfBibitem
\bibitem{Kiselev:2002vz}
V.~V. Kiselev, \ifthenelse{\boolean{articletitles}}{\emph{{Exclusive decays and
  lifetime of \Bcp\ meson in QCD sum rules}},
  }{}\href{http://arxiv.org/abs/hep-ph/0211021}{{\normalfont\ttfamily
  arXiv:hep-ph/0211021}}\relax
\mciteBstWouldAddEndPuncttrue
\mciteSetBstMidEndSepPunct{\mcitedefaultmidpunct}
{\mcitedefaultendpunct}{\mcitedefaultseppunct}\relax
\EndOfBibitem
\bibitem{Ivanov:2006ni}
M.~A. Ivanov, J.~G. Korner, and P.~Santorelli,
  \ifthenelse{\boolean{articletitles}}{\emph{{Exclusive semileptonic and
  nonleptonic decays of the $B_c$ meson}},
  }{}\href{http://dx.doi.org/10.1103/PhysRevD.73.054024}{Phys.\ Rev.\
  \textbf{D73} (2006) 054024},
  \href{http://arxiv.org/abs/hep-ph/0602050}{{\normalfont\ttfamily
  arXiv:hep-ph/0602050}}\relax
\mciteBstWouldAddEndPuncttrue
\mciteSetBstMidEndSepPunct{\mcitedefaultmidpunct}
{\mcitedefaultendpunct}{\mcitedefaultseppunct}\relax
\EndOfBibitem
\bibitem{Hernandez:2006gt}
E.~Hern\'{a}ndez, J.~Nieves, and J.~M. Verde-Velasco,
  \ifthenelse{\boolean{articletitles}}{\emph{{Study of exclusive semileptonic
  and non-leptonic decays of \Bcm\ in a nonrelativistic quark model}},
  }{}\href{http://dx.doi.org/10.1103/PhysRevD.74.074008}{Phys.\ Rev.\
  \textbf{D74} (2006) 074008},
  \href{http://arxiv.org/abs/hep-ph/0607150}{{\normalfont\ttfamily
  arXiv:hep-ph/0607150}}\relax
\mciteBstWouldAddEndPuncttrue
\mciteSetBstMidEndSepPunct{\mcitedefaultmidpunct}
{\mcitedefaultendpunct}{\mcitedefaultseppunct}\relax
\EndOfBibitem
\bibitem{Alves:2008zz}
LHCb collaboration, A.~A. Alves~Jr.\ {\em et~al.},
  \ifthenelse{\boolean{articletitles}}{\emph{{The \lhcb detector at the LHC}},
  }{}\href{http://dx.doi.org/10.1088/1748-0221/3/08/S08005}{JINST \textbf{3}
  (2008) S08005}\relax
\mciteBstWouldAddEndPuncttrue
\mciteSetBstMidEndSepPunct{\mcitedefaultmidpunct}
{\mcitedefaultendpunct}{\mcitedefaultseppunct}\relax
\EndOfBibitem
\bibitem{LHCb-DP-2014-002}
LHCb collaboration, R.~Aaij {\em et~al.},
  \ifthenelse{\boolean{articletitles}}{\emph{{LHCb detector performance}},
  }{}\href{http://dx.doi.org/10.1142/S0217751X15300227}{Int.\ J.\ Mod.\ Phys.\
  \textbf{A30} (2015) 1530022},
  \href{http://arxiv.org/abs/1412.6352}{{\normalfont\ttfamily
  arXiv:1412.6352}}\relax
\mciteBstWouldAddEndPuncttrue
\mciteSetBstMidEndSepPunct{\mcitedefaultmidpunct}
{\mcitedefaultendpunct}{\mcitedefaultseppunct}\relax
\EndOfBibitem
\bibitem{LHCb-DP-2012-002}
A.~A. Alves~Jr.\ {\em et~al.},
  \ifthenelse{\boolean{articletitles}}{\emph{{Performance of the LHCb muon
  system}}, }{}\href{http://dx.doi.org/10.1088/1748-0221/8/02/P02022}{JINST
  \textbf{8} (2013) P02022},
  \href{http://arxiv.org/abs/1211.1346}{{\normalfont\ttfamily
  arXiv:1211.1346}}\relax
\mciteBstWouldAddEndPuncttrue
\mciteSetBstMidEndSepPunct{\mcitedefaultmidpunct}
{\mcitedefaultendpunct}{\mcitedefaultseppunct}\relax
\EndOfBibitem
\bibitem{LHCb-DP-2012-004}
R.~Aaij {\em et~al.}, \ifthenelse{\boolean{articletitles}}{\emph{{The \lhcb
  trigger and its performance in 2011}},
  }{}\href{http://dx.doi.org/10.1088/1748-0221/8/04/P04022}{JINST \textbf{8}
  (2013) P04022}, \href{http://arxiv.org/abs/1211.3055}{{\normalfont\ttfamily
  arXiv:1211.3055}}\relax
\mciteBstWouldAddEndPuncttrue
\mciteSetBstMidEndSepPunct{\mcitedefaultmidpunct}
{\mcitedefaultendpunct}{\mcitedefaultseppunct}\relax
\EndOfBibitem
\bibitem{Sjostrand:2006za}
T.~Sj\"{o}strand, S.~Mrenna, and P.~Skands,
  \ifthenelse{\boolean{articletitles}}{\emph{{PYTHIA 6.4 physics and manual}},
  }{}\href{http://dx.doi.org/10.1088/1126-6708/2006/05/026}{JHEP \textbf{05}
  (2006) 026}, \href{http://arxiv.org/abs/hep-ph/0603175}{{\normalfont\ttfamily
  arXiv:hep-ph/0603175}}\relax
\mciteBstWouldAddEndPuncttrue
\mciteSetBstMidEndSepPunct{\mcitedefaultmidpunct}
{\mcitedefaultendpunct}{\mcitedefaultseppunct}\relax
\EndOfBibitem
\bibitem{Sjostrand:2007gs}
T.~Sj\"{o}strand, S.~Mrenna, and P.~Skands,
  \ifthenelse{\boolean{articletitles}}{\emph{{A brief introduction to PYTHIA
  8.1}}, }{}\href{http://dx.doi.org/10.1016/j.cpc.2008.01.036}{Comput.\ Phys.\
  Commun.\  \textbf{178} (2008) 852},
  \href{http://arxiv.org/abs/0710.3820}{{\normalfont\ttfamily
  arXiv:0710.3820}}\relax
\mciteBstWouldAddEndPuncttrue
\mciteSetBstMidEndSepPunct{\mcitedefaultmidpunct}
{\mcitedefaultendpunct}{\mcitedefaultseppunct}\relax
\EndOfBibitem
\bibitem{LHCb-PROC-2010-056}
I.~Belyaev {\em et~al.}, \ifthenelse{\boolean{articletitles}}{\emph{{Handling
  of the generation of primary events in Gauss, the LHCb simulation
  framework}}, }{}\href{http://dx.doi.org/10.1088/1742-6596/331/3/032047}{{J.\
  Phys.\ Conf.\ Ser.\ } \textbf{331} (2011) 032047}\relax
\mciteBstWouldAddEndPuncttrue
\mciteSetBstMidEndSepPunct{\mcitedefaultmidpunct}
{\mcitedefaultendpunct}{\mcitedefaultseppunct}\relax
\EndOfBibitem
\bibitem{Chang:2005hq}
C.-H. Chang, J.-X. Wang, and X.-G. Wu,
  \ifthenelse{\boolean{articletitles}}{\emph{{BCVEGPY2.0: A Upgrade version of
  the generator BCVEGPY with an addendum about hadroproduction of the P-wave
  B(c) states}}, }{}\href{http://dx.doi.org/10.1016/j.cpc.2005.09.008}{Comput.\
  Phys.\ Commun.\  \textbf{174} (2006) 241},
  \href{http://arxiv.org/abs/hep-ph/0504017}{{\normalfont\ttfamily
  arXiv:hep-ph/0504017}}\relax
\mciteBstWouldAddEndPuncttrue
\mciteSetBstMidEndSepPunct{\mcitedefaultmidpunct}
{\mcitedefaultendpunct}{\mcitedefaultseppunct}\relax
\EndOfBibitem
\bibitem{Allison:2006ve}
Geant4 collaboration, J.~Allison {\em et~al.},
  \ifthenelse{\boolean{articletitles}}{\emph{{Geant4 developments and
  applications}}, }{}\href{http://dx.doi.org/10.1109/TNS.2006.869826}{IEEE
  Trans.\ Nucl.\ Sci.\  \textbf{53} (2006) 270}\relax
\mciteBstWouldAddEndPuncttrue
\mciteSetBstMidEndSepPunct{\mcitedefaultmidpunct}
{\mcitedefaultendpunct}{\mcitedefaultseppunct}\relax
\EndOfBibitem
\bibitem{Agostinelli:2002hh}
Geant4 collaboration, S.~Agostinelli {\em et~al.},
  \ifthenelse{\boolean{articletitles}}{\emph{{Geant4: A simulation toolkit}},
  }{}\href{http://dx.doi.org/10.1016/S0168-9002(03)01368-8}{Nucl.\ Instrum.\
  Meth.\  \textbf{A506} (2003) 250}\relax
\mciteBstWouldAddEndPuncttrue
\mciteSetBstMidEndSepPunct{\mcitedefaultmidpunct}
{\mcitedefaultendpunct}{\mcitedefaultseppunct}\relax
\EndOfBibitem
\bibitem{PDG2017}
Particle Data Group, C.~Patrignani {\em et~al.},
  \ifthenelse{\boolean{articletitles}}{\emph{{\href{http://pdg.lbl.gov/}{Review
  of particle physics}}},
  }{}\href{http://dx.doi.org/10.1088/1674-1137/40/10/100001}{Chin.\ Phys.\
  \textbf{C40} (2016) 100001}, {and 2017 update}\relax
\mciteBstWouldAddEndPuncttrue
\mciteSetBstMidEndSepPunct{\mcitedefaultmidpunct}
{\mcitedefaultendpunct}{\mcitedefaultseppunct}\relax
\EndOfBibitem
\bibitem{Rogozhnikov:2016bdp}
A.~Rogozhnikov, \ifthenelse{\boolean{articletitles}}{\emph{{Reweighting with
  Boosted Decision Trees}},
  }{}\href{http://dx.doi.org/10.1088/1742-6596/762/1/012036}{J.\ Phys.\ Conf.\
  Ser.\  \textbf{762} (2016), no.~1 012036},
  \href{http://arxiv.org/abs/1608.05806}{{\normalfont\ttfamily
  arXiv:1608.05806}}\relax
\mciteBstWouldAddEndPuncttrue
\mciteSetBstMidEndSepPunct{\mcitedefaultmidpunct}
{\mcitedefaultendpunct}{\mcitedefaultseppunct}\relax
\EndOfBibitem
\bibitem{Ebert:2003cz}
D.~Ebert, R.~N. Faustov, and V.~O. Galkin,
  \ifthenelse{\boolean{articletitles}}{\emph{{Weak decays of the $B_c$ meson to
  charmonium and $D$ mesons in the relativistic quark model}},
  }{}\href{http://dx.doi.org/10.1103/PhysRevD.68.094020}{Phys.\ Rev.\
  \textbf{D68} (2003) 094020},
  \href{http://arxiv.org/abs/hep-ph/0306306}{{\normalfont\ttfamily
  arXiv:hep-ph/0306306}}\relax
\mciteBstWouldAddEndPuncttrue
\mciteSetBstMidEndSepPunct{\mcitedefaultmidpunct}
{\mcitedefaultendpunct}{\mcitedefaultseppunct}\relax
\EndOfBibitem
\bibitem{Bourrely:2008dg}
C.~Bourrely, I.~Caprini, and L.~Lellouch,
  \ifthenelse{\boolean{articletitles}}{\emph{{Model-independent description of
  $B \to \pi \ell \nu$ decays and a determination of $|V_{ub}|$}},
  }{}\href{http://dx.doi.org/10.1103/PhysRevD.79.013008}{Phys.\ Rev.\
  \textbf{D79} (2009) 013008}, Erratum
  \href{http://dx.doi.org/10.1103/PhysRevD.82.099902}{ibid.\   \textbf{D82}
  (2010) 099902}, \href{http://arxiv.org/abs/0807.2722}{{\normalfont\ttfamily
  arXiv:0807.2722}}\relax
\mciteBstWouldAddEndPuncttrue
\mciteSetBstMidEndSepPunct{\mcitedefaultmidpunct}
{\mcitedefaultendpunct}{\mcitedefaultseppunct}\relax
\EndOfBibitem
\bibitem{PhysRevD.79.114018}
X.~X. Wang, W.~Wang, and C.~D. L\"u,
  \ifthenelse{\boolean{articletitles}}{\emph{${B}_c$ to p-wave charmonia
  transitions in covariant light-front approach},
  }{}\href{http://dx.doi.org/10.1103/PhysRevD.79.114018}{Phys.\ Rev.\
  \textbf{D79} (2009) 114018},
  \href{http://arxiv.org/abs/0901.1934}{{\normalfont\ttfamily
  arXiv:0901.1934}}\relax
\mciteBstWouldAddEndPuncttrue
\mciteSetBstMidEndSepPunct{\mcitedefaultmidpunct}
{\mcitedefaultendpunct}{\mcitedefaultseppunct}\relax
\EndOfBibitem
\bibitem{LHCb-PAPER-2013-010}
LHCb collaboration, R.~Aaij {\em et~al.},
  \ifthenelse{\boolean{articletitles}}{\emph{{Observation of $\Bcp\to\jpsi\Dsp$
  and $\Bcp\to\jpsi\Dssp$ decays}},
  }{}\href{http://dx.doi.org/10.1103/PhysRevD.87.112012}{Phys.\ Rev.\
  \textbf{D87} (2013) 112012},
  \href{http://arxiv.org/abs/1304.4530}{{\normalfont\ttfamily
  arXiv:1304.4530}}\relax
\mciteBstWouldAddEndPuncttrue
\mciteSetBstMidEndSepPunct{\mcitedefaultmidpunct}
{\mcitedefaultendpunct}{\mcitedefaultseppunct}\relax
\EndOfBibitem
\bibitem{Skwarnicki:1986xj}
T.~Skwarnicki, {\em {A study of the radiative cascade transitions between the
  Upsilon-prime and Upsilon resonances}}, PhD thesis, Institute of Nuclear
  Physics, Krakow, 1986,
  {\href{http://inspirehep.net/record/230779/}{DESY-F31-86-02}}\relax
\mciteBstWouldAddEndPuncttrue
\mciteSetBstMidEndSepPunct{\mcitedefaultmidpunct}
{\mcitedefaultendpunct}{\mcitedefaultseppunct}\relax
\EndOfBibitem
\bibitem{Cranmer:2000du}
K.~S. Cranmer, \ifthenelse{\boolean{articletitles}}{\emph{{Kernel estimation in
  high-energy physics}},
  }{}\href{http://dx.doi.org/10.1016/S0010-4655(00)00243-5}{Comput.\ Phys.\
  Commun.\  \textbf{136} (2001) 198},
  \href{http://arxiv.org/abs/hep-ex/0011057}{{\normalfont\ttfamily
  arXiv:hep-ex/0011057}}\relax
\mciteBstWouldAddEndPuncttrue
\mciteSetBstMidEndSepPunct{\mcitedefaultmidpunct}
{\mcitedefaultendpunct}{\mcitedefaultseppunct}\relax
\EndOfBibitem
\bibitem{Barlow:1993dm}
R.~J. Barlow and C.~Beeston,
  \ifthenelse{\boolean{articletitles}}{\emph{{Fitting using finite Monte Carlo
  samples}}, }{}\href{http://dx.doi.org/10.1016/0010-4655(93)90005-W}{Comput.\
  Phys.\ Commun.\  \textbf{77} (1993) 219}\relax
\mciteBstWouldAddEndPuncttrue
\mciteSetBstMidEndSepPunct{\mcitedefaultmidpunct}
{\mcitedefaultendpunct}{\mcitedefaultseppunct}\relax
\EndOfBibitem
\bibitem{CERN-OPEN-2012-016}
ROOT collaboration, K.~Cranmer {\em et~al.},
  \ifthenelse{\boolean{articletitles}}{\emph{{HistFactory: A tool for creating
  statistical models for use with RooFit and RooStats}}, }{} Tech. Rep.
  CERN-OPEN-2012-016, Jan, 2012\relax
\mciteBstWouldAddEndPuncttrue
\mciteSetBstMidEndSepPunct{\mcitedefaultmidpunct}
{\mcitedefaultendpunct}{\mcitedefaultseppunct}\relax
\EndOfBibitem
\end{mcitethebibliography}

\newpage

 
\clearpage
\centerline{\large\bf LHCb collaboration}
\begin{flushleft}
\small
R.~Aaij$^{40}$,
B.~Adeva$^{39}$,
M.~Adinolfi$^{48}$,
Z.~Ajaltouni$^{5}$,
S.~Akar$^{59}$,
J.~Albrecht$^{10}$,
F.~Alessio$^{40}$,
M.~Alexander$^{53}$,
A.~Alfonso~Albero$^{38}$,
S.~Ali$^{43}$
% ...

{\footnotesize \it
% ...
$ ^{5}$Clermont Universit{\'e}, Universit{\'e} Blaise Pascal, CNRS/IN2P3, LPC, Clermont-Ferrand, France\\
% ...
$ ^{10}$Fakult{\"a}t Physik, Technische Universit{\"a}t Dortmund, Dortmund, Germany\\
% ...
$ ^{38}$ICCUB, Universitat de Barcelona, Barcelona, Spain\\
$ ^{39}$Universidad de Santiago de Compostela, Santiago de Compostela, Spain\\
$ ^{40}$European Organization for Nuclear Research (CERN), Geneva, Switzerland\\
% ...
$ ^{43}$Nikhef National Institute for Subatomic Physics, Amsterdam, The Netherlands\\
% ...
$ ^{48}$H.H. Wills Physics Laboratory, University of Bristol, Bristol, United Kingdom\\
% ...
$ ^{53}$School of Physics and Astronomy, University of Glasgow, Glasgow, United Kingdom\\
% ...
$ ^{59}$University of Cincinnati, Cincinnati, OH, United States\\
% ...
}
\end{flushleft}

\end{document}